\documentclass{iaus}
\usepackage{graphicx}
\usepackage{epsf}
\usepackage{lscape}
\usepackage{graphicx}

\title[The formation of brown dwarfs]
{The formation of brown dwarfs in discs: Physics, numerics, and observations}   
\author[Dimitris Stamatellos \& Anthony Whitworth]{Dimitris Stamatellos \& Anthony Whitworth}   
\affiliation{School of Physics \& Astronomy, Cardiff University, 5 The Parade, Cardiff, CF24 3AA, UK} 

\pubyear{2010}
\volume{270}  
\pagerange{1-30}
\jname{Computational Star Formation}
\editors{}
\begin{document}

\maketitle

\def\sun{\hbox{$\odot$}}
\def\aj{AJ}%
\def\actaa{Acta Astron.}%
\def\araa{ARA\&A}%
\def\apj{ApJ}%
\def\apjl{ApJ}%
\def\apjs{ApJS}%
          \def\aap{A\&A}%
\def\aapr{A\&A~Rev.}%
\def\aaps{A\&AS}%
\def\mnras{MNRAS}%

\begin{abstract}
A large fraction of brown dwarfs and low-mass stars may form by gravitational fragmentation of relatively massive (a few 0.1 M$_{\sun}$) and extended (a few hundred AU) discs around Sun-like stars. We present an ensemble of radiative hydrodynamic simulations that examine the conditions for disc
fragmentation. We demonstrate that this model can explain the low-mass IMF, the brown dwarf desert, and the binary properties of low-mass stars and brown dwarfs. Observing discs that are undergoing fragmentation is possible but very improbable, as the process of disc fragmentation is short lived (discs fragment within a few thousand years).
\end{abstract}

\firstsection 

\section{Introduction}
\label{intro}
The formation of brown dwarfs (BDs)  and low-mass stars is not well understood (Whitworth et al. 2007). Low-mass objects are difficult to form by gravitational fragmentation of unstable gas, as for masses in the BD regime ($\stackrel{<}{_\sim}~80~{\rm M}_{\rm J}$, where ${\rm M}_{\rm J}$ is the mass of Jupiter), a high density ($\stackrel{>}{_\sim}10^{-16} {\rm g\ cm}^{-3}$) is required for the gas to be Jeans unstable. Padoan \& Nordland (2004)  and Hennebelle \&  Chabrier (2008) suggest that these high density cores can be formed by colliding flows in a turbulent magnetic medium. However, this model requires a large amount of turbulence, and has difficulty in explaining the binary properties of BDs. Additionally, the large number of brown-dwarf mass cores that the theory predicts  have not been observed. Another way to reach the high densities required for the formation of BDs and low-mass stars is in gravitationally unstable discs (Whitworth \& Stamatellos 2006; Stamatellos et al. 2007b; Stamatellos \& Whitworth 2008, 2009a,b). These discs form around newly born stars and grow quickly in mass by accreting material from the infalling envelope. They become unstable if the mass accreted onto them cannot efficiently redistribute its angular momentum outwards in order to accrete onto the central star (Attwood et al. 2009).  BDs  are also thought to form as ejected embryos from star forming regions, i.e. as a by-product of the star formation process (Reipurth \& Clarke 2001).  In this model BDs form the same way as low-mass stars, i.e. in collapsing molecular cores,  but shortly after their formation they are ejected from their parental core and stop accreting any further material. Hence, they do not realise their potential to become hydrogen-burning stars. In this paper we focus on the mechanism of BD  formation by fragmentation of gravitationally unstable discs.

\section{The fragmentation of gravitationally unstable discs}
\label{sec:1}

Discs can fragment if (i) they are massive enough so that gravity overcomes thermal and local centrifugal support (Toomre 1964), and (ii) they can cool fast enough  so that the energy provided by the collapse of a proto-fragment is radiated away and the growth of the proto-fragment continues (Gammie 2001; Rice et al. 2005). Analytical and numerical studies have shown that the cooling time must be on the order of the dynamical time which happens to be similar to the orbital time. 

 Massive (a few times 0.1 M$_{\sun}$), extended ($>100$ AU) discs exist (e.g. Enoch et al. 2009; Jogersen et al. 2005; 2009; Greaves et al. 2008; Andrews et al. 2009; Duchene this volume). We have performed an ensemble of radiative hydrodynamical simulations of such discs. We  assume a star-disc system in which the central  star  has initial mass $M_1=0.7\,{\rm M}_{\sun}$. Initially the disc has mass $M_{_{\rm D}}=0.7\,{\rm M}_{\sun}$, inner radius $R_{_{\rm IN}}=40\,{\rm AU}$, outer radius $R_{_{\rm OUT}}=400\,{\rm AU}$, surface density $\Sigma (R) \propto R^{-7/4}$, temperature $T(R)\propto R^{-1/2}$,  and hence approximately uniform initial Toomre parameter $Q\sim 0.9$. Thus, the disc is at the outset marginally gravitationally unstable. The only parameter that is different between different runs is the random noise from which the gravitational instabilities grow in the disc. The evolution of the disc is initially followed using SPH, until $\sim 70\%$ of the disc mass has been accreted, either onto the stars condensing out of the disc, or onto the central  star; this typically happens within 20  kyr. Then the residual gas is ignored and the long term dynamical evolution of the system is followed up to 200 kyr, using an N-body code. The energy equation and associated radiative transfer are treated with the method of Stamatellos et al. (2007a). The intrinsic radiation of the central star is taken into account but its accretion luminosity is ignored (Stamatellos \& Whitworth 2009a). The radius of the sink representing the star is set to 1~AU. The gravitational instabilities develop quickly and the disc fragments within a few thousand years. Typically 5-10 objects form in each disc, most of them  BDs. The final status of these objects is determined by subsequent accretion of material onto them and by their mutual interactions.

\begin{figure}
\centerline{
\includegraphics[angle=-90,width=0.46\columnwidth]{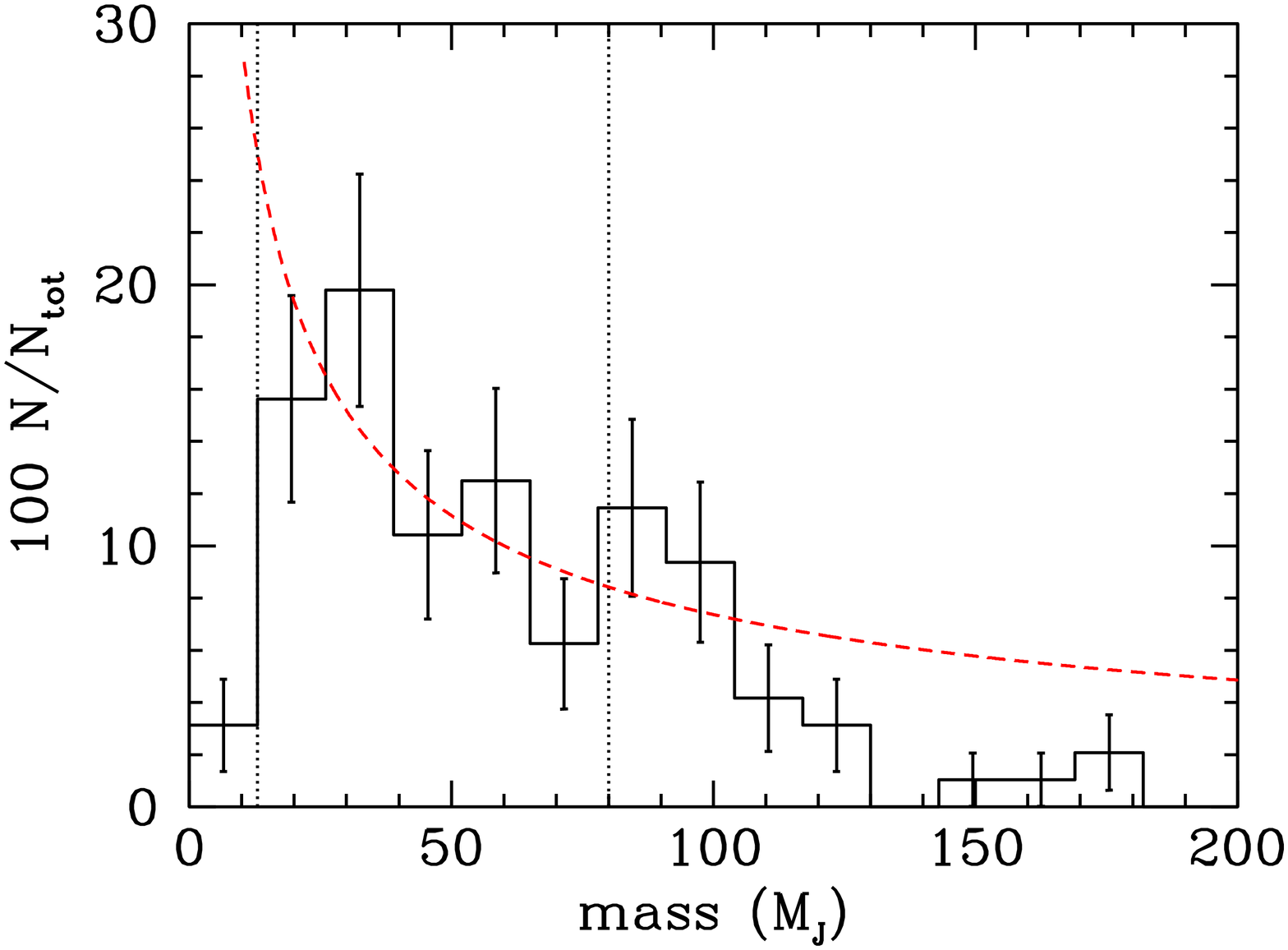}
\includegraphics[angle=-90,width=0.46\columnwidth]{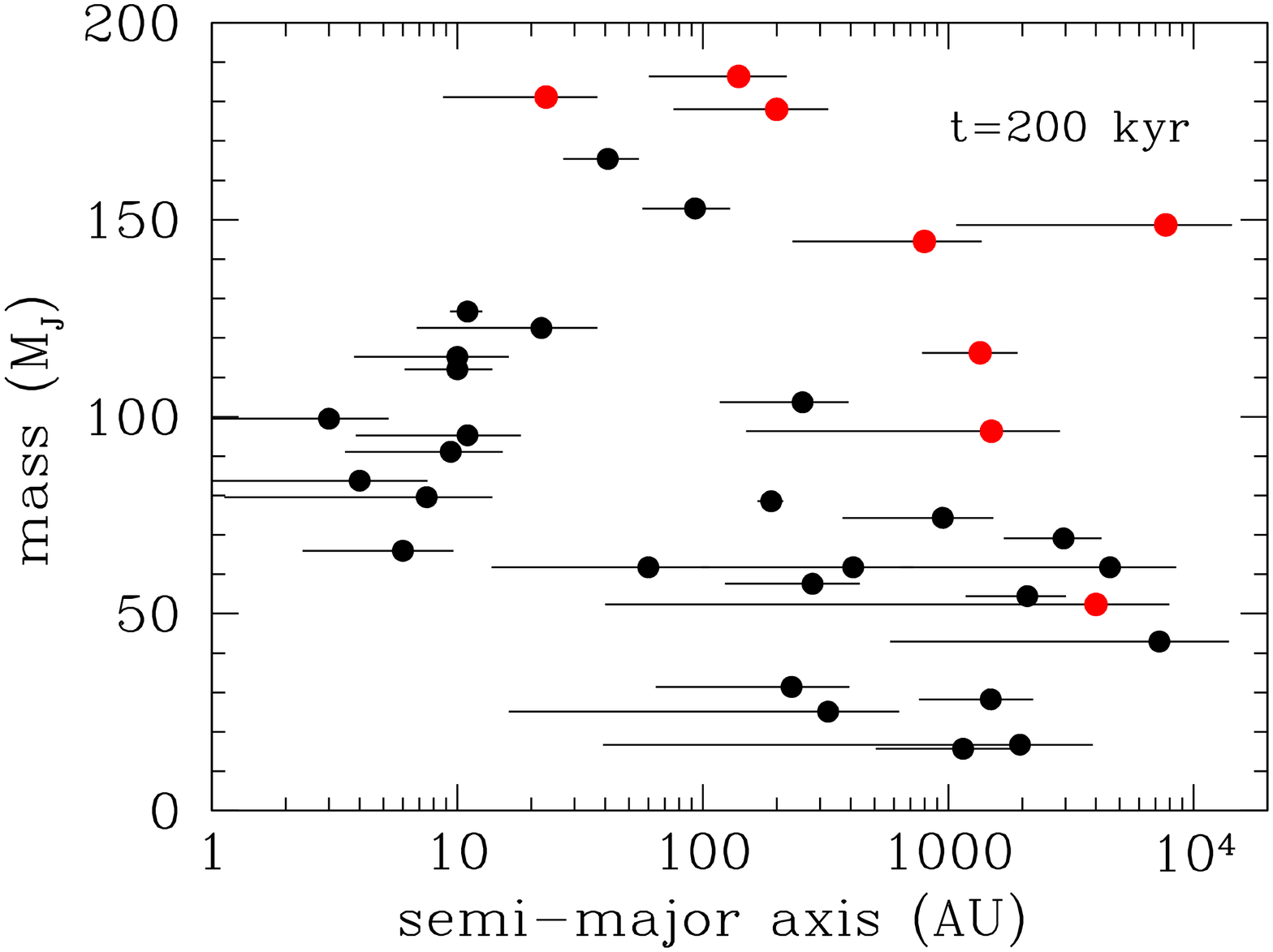}
}
\caption{Left: The mass spectrum of the objects produced by disc fragmentation. Most of these are BDs ($70\%$); the rest are low-mass stars. The vertical dotted lines correspond to the D-burning limit ($\sim 13~{\rm M}_{\rm J}$) and the H-burning limit ($\sim 80~{\rm M}_{\rm J}$). The red dashed line refers to a low-mass IMF  with $\Delta N/\Delta M\propto M^{-0.6}$ . Right: The BD desert. The semi-major axes of the objects (at $t=200$~kyr) formed in the disc and remained bound to the central star plotted against their mass. The bars indicate the minimum and maximum extent of the orbit. There is a lack of BD companions close to central star, but there is a population of BDs loosely bound to the central star. The red dots correspond to low-mass binary systems formed in the disc.}
\label{fig:mspec}   
\label{fig:desert}        
\end{figure}

\section{The statistical properties of objects formed by disc fragmentation}

{\bf The mass spectrum.} Most of the objects  ($\sim 70\%$) formed in the discs are BDs, including a few planetary-mass BDs. The rest are low-mass hydrogen-burning stars. The  typical mass of an object produced is $\sim 20-30~{\rm M}_{\rm J}$. The shape of the initial mass function (Fig.~\ref{fig:mspec}, left) is similar to what observations suggest, i.e. it is consistent with  $\Delta N/\Delta M\propto M^{-\alpha}$, where $\alpha\approx 0.6$ (e.g. in Pleiades  $\alpha=0.6\pm0.11$ -- Moraux et al. 2003 -- and in $\sigma$ Orionis $\alpha=0.6\pm0.1$ -- Lodieu et al. 2009).

{\bf The brown dwarf desert.} Most of the fragments form in the discs with initial masses as low as 3 M$_{\rm J}$ (Stamatellos \& Whitworth 2009c). After a fragment forms it accretes mass from the disc, and interacts with the disc through drag forces, and dynamically with other fragments. As a result some fragments migrate close to the central star. This region is rich in gas, hence these accreting proto-fragments  eventually become low-mass hydrogen-burning stars. Fragments that form farther out also accrete material from the disc but not quite as much; these become BDs. If any of the BDs happen to migrate in the region close to the central star they tend to be ejected back into the outer region through 3-body interactions. Hence, the disc fragmentation model  produces a lack of BD close companions to Sun-like stars, i.e the BD desert (Fig.~\ref{fig:desert}, right). The BD desert may extend out to 300 AU but it is less dry outside 100 AU. We predict a population of BDs at distances from 20 to 5000 AU from the central star. 

{\bf Free-floating planetary-mass objects.} The disc fragmentation model provides an explanation of the existence of free-floating planetary-mass objects (Lucas \& Roche 2000; Zapatero Osorio et al. 2000). In our model these objects form in the disc by gravitational fragmentation and quickly after their formation they are ejected from the system due to 3-body interactions. Hence, they stop accreting and their mass remains low. 

{\bf Low-mass binary statistics.} Close and wide BD-BD binaries are common outcomes of disc fragmentation. The components of the binary form independently in the disc and then pair up. The simulations produce  all kinds of low-mass binaries:  star-star, star-BD, BD-BD, and BD-planetary mass object binaries.  Our model predicts a low-mass binary fraction of $16\%$. This is comparable with the low-mass binary fraction observed in star-forming regions (e.g. in Taurus : $> 20\%$, Kraus et al. 2006, in the field $15\pm5 \%$, Gizis et al. 2003).  We also predict that close low-mass binaries should outnumber wide ones, and this seems to be what is observed (Burgasser et al. 2007).  Most of the low-mass binaries ($55\%$) have components with similar masses ($q>0.7$), in agreement with the observed properties of low-mass binaries (e.g. Burgasser et al. 2007). The model also predicts that BDs that are companions to Sun-like stars are more likely to be in binaries (binary frequency 25\%) than BDs in the field (frequency $5-8\%$). This trend is consistent with what is observed (Burgasser et al. 2005; Faherty et al. 2009).

\section{Numerical Issues}

 We use $1.5\times10^5$ particles to represent each disc, which means that the minimum resolvable mass (corresponding to the number of neighbours used, i.e. 50 SPH particles) is $\approx 0.25\,{\rm M}_{_{\rm J}}$. We have also performed simulations using $2.5\times10^5$  and $4\times10^5$  particles. In these simulations the growth of gravitational instabilities, and the properties of the proto-fragments formed as a result of these instabilities, follow the same patterns as in the simulation with lower resolution, but  the details of the final outcomes (i.e. the number of objects formed and their exact formation positions) are different.

The minimum Jeans mass ($M_{_{\rm J}}=({4\pi^{5/2}}/{24})({c_s^3}/{(G^3\rho)^{1/2}}))$ in a typical simulation using $1.5\times10^5$ particles is $M_{_{\rm JEANS,MIN}}\approx 2~{\rm M}_{\rm J}$. Thus, according to  the Bate \& Burkert (1997) condition, this mass is adequately resolved  as it corresponds to $\sim8\times N_{_{\rm NEIGH}}$ ( a minimum factor of $2\times N_{_{\rm NEIGH}}$ is recommended).
The Toomre mass ($M_T={\pi c_s^4}/{G^2\Sigma}$) is also adequately resolved. The minimum Toomre mass  in our simulation is $M_{_{\rm TOOMRE, MIN}}\approx 2.5~{\rm M}_{\rm J}$). This mass corresponds to $\sim10\times N_{_{\rm NEIGH}}$ ( a factor of  $6\times N_{_{\rm NEIGH}}$ is recommended; Nelson 2006). Finally, the vertical disc structure is resolved by at least $\sim 5$ smoothing lengths, satisfying the Nelson (2006) condition. This is important as the disc mainly cools in this direction.
 
\section{Observing fragmenting discs}

Typically a few BDs form in each fragmenting disc around a  Sun-like star. Hence, it may be that only  $\sim20\%$ of Sun-like stars need to have large and massive unstable discs to produce a large fraction of the observed BDs.  Assuming that the lifetime of the Class 0 phase is $10^5$ yr and that in the disc fragmentation scenario the disc fragments and therefore dissipates within $10^4$ yr, then the probability of observing a fragmenting disc around a Class 0 object is only 10\%. Then, considering that only  20\% of Sun-like stars may have such unstable discs, the probability of observing such discs is only 2\%. Hence, fragmenting discs should be very difficult to discover (Maury et al. 2009).

\section{Conclusions}

Discs can fragment at distances $\stackrel{>}{_\sim} 100$ AU (Stamatellos \& Whiworth 2009a) from the central star to form predominately brown dwarfs, but also  low-mass hydrogen burning stars and planetary-mass objects. Despite the fact that the model does not include magnetic fields, radiative feedback due to the accretion luminosity from newly formed protostars, and mechanical feedback (i.e. jets), it can reasonably reproduce (i) the shape of the low-mass IMF, (ii) the brown dwarf desert, (iii) the binary properties of low-mass objects and (iv) the formation of free-floating planetary mass objects.


\begin{thebibliography}{}

\bibitem[Andrews et al. 2009]{andrews09} Andrews, S.~M., Wilner, D.~J., Hughes, A.~M., Qi, C., \& Dullemond, C.~P.\ 2009, \apj, 700, 1502 

\bibitem[Attwood et al. 2009]{attwood09} Attwood, R.~E., Goodwin, S.~P., Stamatellos, D., \& Whitworth, A.~P.\ 2009, \aap, 495, 201 

\bibitem[]{burgasser05} Burgasser, A.~J., Kirkpatrick, J.~D., \& Lowrance, P.~J.\ 2005, \aj, 129, 2849 

\bibitem[]{burgasser07} Burgasser, A.~J., Reid, I.~N., Siegler, N., Close, L., Allen, P., Lowrance, P., \& Gizis, J.\ 2007, Protostars and Planets V, 427 

\bibitem[]{faherty09} Faherty, J.~K.,  Burgasser, A.~J., West, A.~A., Bochanski, J.~J., Cruz, K.~L., Shara, M.~M., \& Walter, F.~M.\ 2009, arXiv:0911.1363 

\bibitem[]{gammie} Gammie, C.~F.\ 2001, \apj, 553, 174 

\bibitem[]{gizis03} Gizis, J.~E., Reid, I.~N., Knapp, G.~R., Liebert, J., Kirkpatrick, J.~D., Koerner, D.~W., \& Burgasser, A.~J.\ 2003, \aj, 125, 3302 

\bibitem[]{hennebelle} Hennebelle, P., \& Chabrier, G.\ 2008, \apj, 684, 395 

\bibitem[]{kraus06} Kraus, A.~L., White, R.~J., \& Hillenbrand, L.~A.\ 2006, \apj, 649, 306 

\bibitem[]{lodieu09} Lodieu, N., Zapatero Osorio, M.~R., Rebolo, R., Mart{\'{\i}}n, E.~L., \& Hambly, N.~C.\ 2009, \aap, 505, 1115 

\bibitem[]{lucas00} Lucas, P.~W., \& Roche, P.~F.\ 2000, \mnras, 314, 858 

\bibitem[Maury et  al.(2010)]{2010A&A...512A..40M} Maury, A.~J., et al.\ 2010, \aap, 512, A40 

\bibitem[]{moraux03} Moraux, E., Bouvier, J., Stauffer, J.~R., \& Cuillandre, J.-C.\ 2003, \aap, 400, 891 

\bibitem[]{padoan} Padoan, P., \&  Nordlund, {\AA}.\ 2004, \apj, 617, 559 

\bibitem[]{rei} Reipurth, B., \& Clarke, C.\ 2001, \aj, 122, 432

\bibitem[]{rice} Rice, W.~K.~M., Lodato, G., \& Armitage, P.~J.\ 2005, \mnras, 364, L56 

\bibitem[]{stam07a} Stamatellos, D., Whitworth, A.~P., Bisbas, T., \& Goodwin, S.\ 2007a, \aap, 475, 37 

\bibitem[]{stam07b} Stamatellos, D., Hubber, D.~A., \& Whitworth, A.~P.\ 2007b, \mnras, 382, L30 

\bibitem[]{stam08} Stamatellos, D., \& Whitworth, A.~P.\ 2008, \aap, 480, 879 

\bibitem[]{stam09a} Stamatellos, D., \& Whitworth, A.~P.\ 2009a, \mnras, 392, 413 

\bibitem[]{stam09c} Stamatellos, D., \& Whitworth, A.~P.\ 2009b, \mnras, 1548


\bibitem[]{toomre} Toomre, A.\ 1964, \apj, 139, 1217

\bibitem[]{whit06} Whitworth, A.~P., \& Stamatellos, D.\ 2006, \aap, 458, 817 

\bibitem[]{whit2007} Whitworth, A., Bate, M.~R., Nordlund, {\AA}., Reipurth, B., \& Zinnecker, H.\ 2007, Protostars and Planets V, 459 

\bibitem[] {zapatero00} Zapatero Osorio, M.~R., B{\'e}jar, V.~J.~S., Mart{\'{\i}}n, E.~L., Rebolo, R., y  Navascu{\'e}s, D.~B., Bailer-Jones, C.~A.~L., \& Mundt, R.\ 2000, Science, 290, 103 



\end{thebibliography}
\end{document}